\documentclass[12pt,prd,showpacs,tightenlines,nofootinbib]{revtex4}
\usepackage{bm}
\usepackage{graphics}
\usepackage{rotating}
\usepackage{epsfig}
\begin{document}
\begin{flushright}{HU-EP-11/47}\end{flushright}
\title{Spectroscopy and Regge trajectories of heavy quarkonia and
$B_c$ mesons}
\author{D. Ebert$^{1}$, R. N. Faustov$^{2}$  and V. O. Galkin$^{1,2}$}
\affiliation{
$^1$ Institut f\"ur Physik, Humboldt--Universit\"at zu Berlin,
Newtonstr. 15, D-12489  Berlin, Germany\\
$^2$ Dorodnicyn Computing Centre, Russian Academy of Sciences,
  Vavilov Str. 40, 119991 Moscow, Russia}

\begin{abstract}
The mass spectra of charmonia, bottomonia and $B_c$ mesons are
calculated in the framework of the QCD-motivated relativistic quark
model based on the quasipotential approach. The dynamics of heavy
quarks and antiquarks is treated fully relativistically without
application of the nonrelativistic $v^2/c^2$ expansion. The known
one-loop radiative corrections  to the heavy quark potential are taken
into account perturbatively. The heavy quarkonium masses are
calculated  up to rather high orbital and radial excitations ($L=5$, $n_r=5$). On
this basis the Regge trajectories are constructed both in the total
angular momentum $J$ and radial quantum number $n_r$. It is found that
the daughter trajectories are almost linear and parallel, while parent
trajectories exhibit some nonlinearity in the low mass region. Such
nonlinearity is most pronounced for bottomonia and is only marginal
for charmonia. The obtained results are compared  with the available
experimental data, and a possible interpretation of the new
charmonium-like states above open charm production threshold is discussed.   

\end{abstract}


\maketitle

\section{Introduction}
\label{sec:intr}
At present a vast amount of experimental data on the heavy quarkonium
spectroscopy has been accumulated \cite{pdg}. The number of known states is
constantly increasing. Thus, in the last eight years more than ten new
charmonium-like states have been discovered (for a recent review see
e.g. \cite{qwg}). The total number of
charmonium states, listed in the Particle Data Group Listings \cite{pdg},
is 25 at present. Some of
the new states (such as $\eta_c(2S)$, $h_c$, $\chi_{c2}(2P)$, etc.) are
the long-awaited ones, expected by quark models many 
years ago, while some others, with masses higher than the
threshold of the open charm production, have narrow widths and
unexpected decay properties \cite{qwg}. There are theoretical indications
that some of these new states could be the first manifestation of the existence of
exotic hadrons (tetraquarks, molecules, hybrids etc.), which are
expected to exist in QCD (see e.g. \cite{kz} and references therein).  In order to explore such options, a
comprehensive understanding of the heavy quarkonium spectroscopy up to
rather high orbital and radial excitations is required. The LHCb
Collaboration at the Large Hadron Collider (LHC) \cite{lhcbp} plans to search for the bottom
counterparts of the newly discovered charmonium-like states \cite{lhcbqq}. At
present, the experimentally known bottomonium spectrum consists of 14
states \cite{pdg}. Moreover, new experimental data on the spectroscopy of $B_c$ mesons are
expected from LHC \cite{lhcb}. Therefore, the investigation
of the  masses of the excited heavy quarkonia states represents an important and
interesting problem. A reliable calculation of the masses of  excited
heavy quark-antiquark
states will allow to single out  experimental candidates for the exotic
multiquark states.

In Ref.~\cite{efg} we calculated the mass spectra of charmonia, bottomonia
and $B_c$ mesons on the basis of a three-dimensional relativistic quasipotential
wave equation with a QCD-motivated potential. For this calculation,
a $v^2/c^2$ expansion up to the second order was used. Masses of several lowest
orbital and radial excitations were obtained, mainly for the states
lying under the open flavour production threshold.  This investigation
indicated that the charm quark is not heavy enough to be considered as
nonrelativistic, especially for excited states \cite{efg}. Therefore, a 
reliable consideration of the highly excited charmonium states requires a
completely relativistic treatment of the charmed quark without an
expansion in its velocity. In this paper we  extend
the approach previously used for the investigations of light meson
spectroscopy \cite{lregge}, where relativistic quark dynamics was
treated completely relativistically,  to heavy quarkonia. Then the
relativistic quasipotential, 
which determines the quark dynamics in heavy quarkonia, 
is an extremely non-local function in the coordinate  
space. In order to make it local, we replace the quark energies,
entering the quark spinors, with the corresponding on-mass-shell
energies. Such procedure makes the quasipotential local, but
introduces a rather complicated nonlinear dependence on the bound state
mass. The quasipotential equation with the complete relativistic
potential can then be solved 
numerically using previously developed numerical methods \cite{lmes}. In order to
improve our description, leading radiative corrections to the heavy quark
potential \cite{radcorr} are also taken into account. Such corrections are
suppressed by additional powers of $\alpha_s$, which are rather small
for heavy quarkonia, and are known only in the framework of the
$v^2/c^2$ expansion. Therefore we treat them perturbatively. The
calculation of the masses of highly orbitally and 
radially excited states up to the fifths excitation is carried out. On this
basis, the Regge trajectories for charmonia, bottomonia and $B_c$
mesons can be constructed both in the total angular momentum $J$ and
radial quantum number $n_r$, and properties like linearity, parallelism and equidistance
of these trajectories can be checked. There are reasons to expect that
the parent Regge trajectories can be nonlinear due to the compactness
of their ground and lowest excited states, which puts them in the
region where both the linear confining and Coulomb  parts of
the quark-antiquark potential play an equally important role. Note
also that the
possibility of the assignment of the experimentally observed highly
excited heavy quarkonium states to a particular Regge trajectory
could help in determining their quantum numbers and elucidating their
nature.  

In recent papers \cite{tetr,htetr} we investigated the possible
interpretation of some new unconventional charmonium-like states
\cite{qwg} as diquark-antidiquark tetraquarks. In particular, the relativistic
dynamical calculation of the masses of such states was performed. Here
we complement this study by calculating the spectrum of the
highly excited  conventional heavy quark-antiquark states in the same mass
region. As a result, we will obtain a consistent picture of the recently
discovered heavy quarkonium states within the same
relativistic quark model.   

A vast literature on the heavy quarkonium spectroscopy is now
available. Therefore, we mostly refer to the recent comprehensive
reviews \cite{qwg,godfrey,swanson,ali},  where the references to earlier
review and original papers can be found.  Recent investigations of
highly excited heavy quarkonium states and their Regge trajectories can be
found, e.g., in Refs.~\cite{gll,wg,badalian}. For very recent
unquenched lattice QCD calculations see, e.g., Ref.~\cite{bali} and
references therein.

\section{Relativistic quark model}
\label{sec:rqm}

  In the relativistic quark model based on the quasipotential approach
  a meson is described by the wave 
function of the bound quark-antiquark state, which satisfies the
quasipotential equation  of the Schr\"odinger type~\cite{efg}
\begin{equation}
\label{quas}
{\left(\frac{b^2(M)}{2\mu_{R}}-\frac{{\bf
p}^2}{2\mu_{R}}\right)\Psi_{M}({\bf p})} =\int\frac{d^3 q}{(2\pi)^3}
 V({\bf p,q};M)\Psi_{M}({\bf q}),
\end{equation}
where the relativistic reduced mass is
\begin{equation}
\mu_{R}=\frac{E_1E_2}{E_1+E_2}=\frac{M^4-(m^2_1-m^2_2)^2}{4M^3},
\end{equation}
and $E_1$, $E_2$ are given by
\begin{equation}
\label{ee}
E_1=\frac{M^2-m_2^2+m_1^2}{2M}, \quad E_2=\frac{M^2-m_1^2+m_2^2}{2M}.
\end{equation}
Here $M=E_1+E_2$ is the meson mass, $m_{1,2}$ are the quark masses,
and ${\bf p}$ is their relative momentum.  
In the center-of-mass system the relative momentum squared on mass shell 
reads
\begin{equation}
{b^2(M) }
=\frac{[M^2-(m_1+m_2)^2][M^2-(m_1-m_2)^2]}{4M^2}.
\end{equation}

The kernel 
$V({\bf p,q};M)$ in Eq.~(\ref{quas}) is the quasipotential operator of
the quark-antiquark interaction. It is constructed with the help of the
off-mass-shell scattering amplitude, projected onto the positive
energy states. 
Constructing the quasipotential of the quark-antiquark interaction, 
we have assumed that the effective
interaction is the sum of the usual one-gluon exchange term with the mixture
of long-range vector and scalar linear confining potentials, where
the vector confining potential
contains the Pauli interaction. The quasipotential is then defined by
  \begin{equation}
\label{qpot}
V({\bf p,q};M)=\bar{u}_1(p)\bar{u}_2(-p){\mathcal V}({\bf p}, {\bf
q};M)u_1(q)u_2(-q),
\end{equation}
with
$${\mathcal V}({\bf p},{\bf q};M)=\frac{4}{3}\alpha_sD_{ \mu\nu}({\bf
k})\gamma_1^{\mu}\gamma_2^{\nu}
+V^V_{\rm conf}({\bf k})\Gamma_1^{\mu}
\Gamma_{2;\mu}+V^S_{\rm conf}({\bf k}),$$
where $\alpha_s$ is the QCD coupling constant, $D_{\mu\nu}$ is the
gluon propagator in the Coulomb gauge,
 $\gamma_{\mu}$ and $u(p)$ are 
the Dirac matrices and spinors and ${\bf k=p-q}$.

The effective long-range vector vertex is
given by
\begin{equation}
\label{kappa}
\Gamma_{\mu}({\bf k})=\gamma_{\mu}+
\frac{i\kappa}{2m}\sigma_{\mu\nu}k^{\nu},
\end{equation}
where $\kappa$ is the Pauli interaction constant characterizing the
anomalous chromomagnetic moment of quarks. Vector and
scalar confining potentials in the nonrelativistic limit reduce to
\begin{eqnarray}
\label{vlin}
V^V_{\rm conf}(r)&=&(1-\varepsilon)(Ar+B),\nonumber\\ 
V^S_{\rm conf}(r)& =&\varepsilon (Ar+B),
\end{eqnarray}
reproducing 
\begin{equation}
\label{nr}
V_{\rm conf}(r)=V^S_{\rm conf}(r)+V^V_{\rm conf}(r)=Ar+B,
\end{equation}
where $\varepsilon$ is the mixing coefficient. Therefore, in this
 limit the Cornell-type potential is reproduced
 \begin{equation}
\label{cp}
V_{\rm NR}(r)=-\frac43\frac{\alpha_s}{r}+Ar+B.
\end{equation}

All the model parameters have the same values as in our previous
papers \cite{hlmass,efg}.
The constituent quark masses $m_u=m_d=0.33$ GeV, $m_s=0.5$ GeV,
$m_c=1.55$ GeV, $m_b=4.88$ GeV, and
the parameters of the linear potential $A=0.18$ GeV$^2$ and $B=-0.16$ GeV
have the usual values of quark models.  The value of the mixing
coefficient of vector and scalar confining potentials $\varepsilon=-1$
has been determined from the consideration of charmonium radiative
decays \cite{efg} and matching heavy quark effective theory (HQET). 
Finally, the universal Pauli interaction constant $\kappa=-1$ has been
fixed from the analysis of the fine splitting of heavy quarkonia ${
}^3P_J$- states \cite{efg}. In this case, the long-range chromomagnetic
interaction of quarks, which is proportional to $(1+\kappa)$, vanishes
in accordance with the flux-tube model.

\section{Quark-antiquark potential}
\label{sec:bc}
The investigations of the heavy quark dynamics in heavy mesons indicate that
the charm quark is not heavy enough to be considered as
nonrelativistic. Indeed, estimates of the averaged velocity squared for
the ground-state charmonium give the value $\langle v^2/c^2\rangle\sim 0.25$. For
excited charmonium states the $\langle v^2/c^2\rangle$ values are even
higher. Therefore, a 
reliable calculation of the charmonium spectroscopy requires a
completely relativistic treatment of the charmed quark without an
expansion in its velocity.

The quasipotential (\ref{qpot})  can in principal  be used for arbitrary quark
masses.  The substitution 
of the Dirac spinors  into (\ref{qpot}) results in an extremely
nonlocal potential in the configuration space. Clearly, it is very hard to 
deal with such potentials without any additional approximations.
 In order to simplify the relativistic $Q\bar Q$ potential, we make the
following replacement in the Dirac spinors:
\begin{equation}
  \label{eq:sub}
  \epsilon_{1,2}(p)=\sqrt{m_{1,2}^2+{\bf p}^2} \to E_{1,2}
\end{equation}
(see the discussion of this point in \cite{hlmass,lmes}).  This substitution
makes the Fourier transformation of the potential (\ref{qpot}) local.

The resulting $Q\bar Q$ potential then reads
\begin{equation}
  \label{eq:v}
  V(r)= V_{\rm SI}(r)+ V_{\rm SD}(r),
\end{equation}
where the explicit expression for the spin-independent $V_{\rm SI}(r)$
can be found in Ref.~\cite{lregge}.
The structure of the spin-dependent potential is given by
\begin{equation}
  \label{eq:vsd}
  V_{\rm SD}(r)= a\ {\bf L}\cdot{\bf S}+b\left[\frac{3}{r^2}({\bf S}_1\cdot
{\bf r})({\bf S}_2\cdot {\bf r})-({\bf S}_1\cdot {\bf S}_2)\right] +c\ {\bf
S}_1\cdot {\bf S}_2 +d\ {\bf L}\cdot({\bf S}_1-{\bf S}_2)+ e\, ({\bf L}{\bf
  S}_1) ({\bf L}{\bf S}_2),
\end{equation}
where $ {\bf L}$ is the orbital angular momentum, ${\bf S}_i$ is the
quark spin,  ${\bf S}={\bf S}_1+ {\bf S}_2$. The coefficients $a$, $b$, $c$, $d$ and $e$ are expressed
through the corresponding derivatives of the Coulomb and
confining potentials. Their explicit expressions are given in
Ref.~\cite{lregge}. Since we also include the one-loop radiative
corrections in our calculations, the strong coupling constant
$\alpha_s$ in the static potential (\ref{cp}) should be replaced by   
the corrected constant $\bar\alpha_V$ \cite{radcorr}:
\begin{eqnarray}\label{alphav}
\bar\alpha_V(\mu^2)&=&\alpha_s(\mu^2)\left[1+\left(\frac{a_1}{4}
+\frac{\gamma_E\beta_0}{2}\right)\frac{\alpha_s(\mu^2)}{\pi}\right],\\
a_1&=&\frac{31}{3}-\frac{10}{9}n_f,\qquad
\beta_0=11-\frac23n_f,\nonumber
\end{eqnarray} 
where  $n_f$ is the number of flavours, $\mu$ is a renormalization
scale and $\gamma_E\cong 0.5772$ is the Euler constant.

The resulting quasipotential equation with the complete kernel
(\ref{eq:v}) is solved numerically without any approximations.
The remaining one-loop radiative corrections, which are not included
in the renormalized coupling constant (\ref{alphav}), are known only
in the framework of the $v/c$ expansion \cite{radcorr}; therefore we  treat them
perturbatively. The additional contributions are the  following \cite{radcorr,efg}: \\
(a) to the spin-independent part
\begin{eqnarray}
\label{sipot}
\delta V_{\rm SI}(r)&=& -\frac43
\frac{\beta_0 \alpha_s^2(\mu^2)}{2\pi}\frac{\ln(\mu r)}{r}
 +\frac18\left(\frac{1}{m_1^2}+\frac{1}{m_2^2}\right) 
\Delta\left[ -\frac43\frac{\beta_0\alpha_s^2(\mu^2)}{2\pi}
\frac{\ln(\mu r)}{r}\right]\cr\nonumber\\
&&+\frac{1}{2m_1m_2}\left[-\frac43\frac{\beta_0\alpha_s^2(\mu^2)}{2\pi}\left\{{\bf p}^2
  \frac{\ln(\mu r)}{r} +\frac{({\bf p\cdot 
r})^2}{r^2}\left(\frac{\ln(\mu r)}{r}-\frac1r\right)\right\}_W\right],
\end{eqnarray}
(b) to the spin-dependent part
\begin{eqnarray}
\label{a}
\delta a&=& \frac14\left(\frac{1}{m_1^2}+\frac{1}{m_2^2}\right)
  \frac43\frac{\alpha_s^2(\mu^2)}{\pi r^3}\Biggl[\frac73-\frac{\beta_0}{12}+
  \gamma_E\left(\frac{\beta_0}{2}-3\right)+\frac{\beta_0}{2}\ln(\mu r)
-3\ln(\sqrt{m_1m_2}\,r)\Biggr]\cr\cr
&&+\frac1{m_1m_2}\frac43\frac{\alpha_s^2(\mu^2)}{\pi r^3}
\Biggl[\frac16-\frac{\beta_0}{12}+\gamma_E\left(\frac{\beta_0}{2}-\frac32\right)+\frac{\beta_0}{2}
\ln(\mu r)
-\frac32\ln(\sqrt{m_1m_2}\,r)\Biggr]\cr\cr
&&+\left(\frac{1}{m_1^2}-\frac{1}{m_2^2}\right)
  \frac{\alpha_s^2(\mu^2)}{2\pi r^3}\ln\frac{m_2}{m_1},\\ \cr
\label{b}
\delta b&=& \frac{1}{3m_1 m_2}\frac{4\alpha_s^2(\mu^2)}{\pi r^3}\Biggl[\frac{29}{6}-\frac{1}{4}\beta_0+
\gamma_E\left(\frac{\beta_0}{2}-3\right)+\frac{\beta_0}{2}\ln(\mu r)
-3\ln(\sqrt{m_1m_2}\,r)\Biggr],\\ \cr
\label{c}
\delta c&=& \frac{4}{3m_1 m_2}\frac{8\pi\alpha_s^2(\mu^2)}{3\pi}\Biggl\{\left(\frac{5}{12}\beta_0-\frac{11}{3} 
-\left[\frac{m_1-m_2}{m_1+m_2}+\frac18\,\frac{m_1+m_2}{m_1-m_2}\right]
\ln\frac{m_2}{m_1}\right)
\delta^3(r)\cr\nonumber\\
&&+\left[-
\frac{\beta_0}{8\pi}\nabla^2\left(\frac{\ln({\mu} r)+\gamma_E}{r}\right)
+\frac{21}{16\pi}\nabla^2\left(
\frac{\ln(\sqrt{m_1m_2}\,r)+\gamma_E}{r}\right)\right]\Biggr\},\\ \cr
\label{d}
\delta d&=& \frac14\left(\frac{1}{m_1^2}-\frac{1}{m_2^2}\right)
  \frac43\frac{\alpha_s^2(\mu^2)}{\pi r^3}\Biggl[\frac73-\frac{\beta_0}{12}+
  \gamma_E\left(\frac{\beta_0}{2}-3\right)+\frac{\beta_0}{2}\ln(\mu r)
-3\ln(\sqrt{m_1m_2}\,r)\Biggr]\cr\cr
&&+\left(\frac{1}{m_1}+\frac{1}{m_2}\right)^2
  \frac{\alpha_s^2(\mu^2)}{2\pi r^3}\ln\frac{m_2}{m_1},
   \end{eqnarray}
where for quantities quadratic in the momenta we use the
Weyl ordering prescription \cite{bcp}:
$$\{f(r)p^ip^j\}_W=\frac14\{\{f(r),p^i\},p^j\}.$$
Since we consider only heavy quarks, for the dependence of the QCD coupling constant $\alpha_s(\mu^2)$ 
on the renormalization point $\mu^2$ we use the leading order
result
\begin{equation}
\label{alpha}
\alpha_s(\mu^2)=\frac{4\pi}{\beta_0\ln(\mu^2/\Lambda^2)}.
\end{equation}
In our numerical calculations we set the renormalization scale 
$\mu=2m_1m_2/(m_1+m_2)$ and $\Lambda=0.169$~GeV, which gives
$\alpha_s=0.315$ for $m_1=m_2=m_c$ (charmonium); $\alpha_s=0.224$ for
$m_1=m_2=m_b$ (bottomonium); and $\alpha_s=0.286$ for $m_1=m_c$, $m_2=m_b$
($B_c$ meson).~\footnote{Note that these values of $\alpha_s(\mu^2)$
  were fixed in Ref.~\cite{efg} to optimize the fine and hyperfine
  splittings in charmonium and bottomonium (cf.
  Tables~\ref{tab:csm}, \ref {tab:bsm}).}

For the equal mass 
case ($m_1=m_2=m$) the contribution of the
annihilation diagrams  
\begin{equation}
  \label{eq:deltc}
  \delta c'=\frac{8\alpha_s^2(\mu^2)}{3m^2}\left(\frac43-\ln
  2\right)\delta^3(r), 
\end{equation}
which is of second order in $\alpha_s$, must be added to the spin-spin interaction coefficient $\delta c$ in
Eq.~(\ref{c}). 

Moreover, for the calculation of the bottomonium mass spectrum it is also
necessary to take into account 
additional one-loop corrections due to the finite mass of the charm quark
\cite{fmc}. We considered these corrections within our model in
Ref.~\cite{efgmc} and found that they give contributions of a few MeV
and weakly depend on the quantum numbers of the bottomonium
states. The one-loop correction to the static $Q\bar Q$ potential in QCD
due to the finite $c$ quark mass is given by \cite{fmc,efgmc}
\begin{equation}
  \label{eq:deltav}
  \Delta V(r,m_c)= -\frac49\frac{\alpha_s^2(\mu)}{\pi
  r}\left[\ln(\sqrt{a_0}m_c r) +\gamma_E+E_1(\sqrt{a_0}m_c
  r)\right],
\end{equation}  
where
\[
E_1(x)=\int_x^\infty e^{-t}\ \frac{dt}t=-\gamma_E-\ln x-
\sum_{n=1}^\infty \frac{(-x)^n}{n\cdot n!},
\] 
 and $a_0=5.2$. 

\section{Results and discussion}
\label{sec:rd}

\subsection{Calculation of the masses}
\label{sec:cm}

We solve the quasipotential equation with the
quasipotential (\ref{eq:v}), which nonperturbatively accounts for the
relativistic dynamics of both heavy quarks,  numerically. Then we add
the one-loop
radiative corrections (\ref{sipot})-(\ref{eq:deltc}) and the additional
one-loop correction for bottomonium due to the finite mass of the
charmed quark (\ref{eq:deltav}) by  using perturbation theory. 
The calculated masses of charmonia, bottomonia and the $B_c$ meson are
given in Tables~\ref{tab:csm}-\ref{tab:bcms}, where $n=n_r+1$, $n_r$
is the radial quantum number, 
$L$, $S$ and $J$ are the quantum numbers of the orbital, spin and
total angular momenta, respectively. They  are confronted
with available experimental data from PDG \cite{pdg}. Good agreement
of our predictions and data is found. It is important to note that the
nonperturbative relativistic treatment gives a better agreement with
data than our previous heavy quarkonium mass spectrum calculation
\cite{efg}, where only relativistic corrections up to $v^2/c^2$ order were
taken into account. However, the differences between former and new
predictions are rather small for most of the low-lying states and become
noticeable only for higher excitations, where relativistic effects turn
out to be particularly important.

\begin{table}
\caption{Charmonium mass spectrum
   (in MeV).} 
   \label{tab:csm}
\begin{ruledtabular}
\begin{tabular}{cccccccccc}
\multicolumn{2}{l}{\underline{\phantom{p}\hspace{0.8cm}State\hspace{0.8cm}}}&Theory
&\multicolumn{2}{l}{\underline{\hspace{0.7cm}Experiment \cite{pdg}\hspace{0.7cm}}}&\multicolumn{2}{l}{\underline{\phantom{p}\hspace{0.8cm}State\hspace{0.8cm}}}&
Theory&\multicolumn{2}{r}{\underline{\hspace{0.7cm}Experiment \cite{pdg}\hspace{0.7cm}}}\\
$n^{2S+1}L_J$&$J^{PC}$& & meson &mass&$n^{2S+1}L_J$&$J^{PC}$&  & meson&mass\\[2pt]
\hline
$1^1S_0$& $0^{-+}$&2981& $\eta_c(1S)$&2980.3(1.2)&$2^3D_1$& $1^{--}$&4150& $\psi(4160)$&4153(3)\\
$1^3S_1$& $1^{--}$&3096& $J/\psi(1S)$&3096.916(11)&$2^3D_2$& $2^{--}$&4190& &\\
$2^1S_0$& $0^{-+}$&3635& $\eta_c(2S)$&3637(4)&$2^3D_3$& $3^{--}$&4220& &\\
$2^3S_1$& $1^{--}$&3685& $\psi(2S)$&3686.09(4)&$2^1D_2$& $2^{-+}$&4196& $X(4160)?$&4156($^{29}_{25}$)\\
$3^1S_0$& $0^{-+}$&3989& &&$3^3D_1$& $1^{--}$&4507& &\\
$3^3S_1$& $1^{--}$&4039& $\psi(4040)$&4039(1)&$3^3D_2$& $2^{--}$&4544& &\\
$4^1S_0$& $0^{-+}$&4401& &&$3^3D_3$& $3^{--}$&4574& &\\
$4^3S_1$& $1^{--}$&4427& $\psi(4415)$&4421(4)&$3^1D_2$& $2^{-+}$&4549& &\\
$5^1S_0$& $0^{-+}$&4811& &&$4^3D_1$& $1^{--}$&4857& &\\
$5^3S_1$& $1^{--}$&4837& &&$4^3D_2$& $2^{--}$&4896& &\\
$6^1S_0$& $0^{-+}$&5155& &&$4^3D_3$& $3^{--}$&4920& &\\
$6^3S_1$& $1^{--}$&5167& &&$4^1D_2$& $2^{-+}$&4898& &\\
$1^3P_0$& $0^{++}$&3413& $\chi_{c0}(1P)$&3414.75(31)&$1^3F_2$& $2^{++}$&4041& &\\
$1^3P_1$& $1^{++}$&3511& $\chi_{c1}(1P)$&3510.66(7)&$1^3F_3$& $3^{++}$&4068& &\\
$1^3P_2$& $2^{++}$&3555& $\chi_{c2}(1P)$&3556.20(9)&$1^3F_4$& $4^{++}$&4093& &\\
$1^1P_1$& $1^{+-}$&3525& $h_{c}(1P)$&3525.41(16)&$1^1F_3$& $3^{+-}$&4071& &\\
$2^3P_0$& $0^{++}$&3870& &&$2^3F_2$& $2^{++}$&4361& &\\
$2^3P_1$& $1^{++}$&3906& &&$2^3F_3$& $3^{++}$&4400& &\\
$2^3P_2$& $2^{++}$&3949& $\chi_{c2}(2P)$&3927.2(2.6)&$2^3F_4$& $4^{++}$&4434& &\\
$2^1P_1$& $1^{+-}$&3926& &&$2^1F_3$& $3^{+-}$&4406& &\\
$3^3P_0$& $0^{++}$&4301& &&$1^3G_3$& $3^{--}$&4321& &\\
$3^3P_1$& $1^{++}$&4319& &&$1^3G_4$& $4^{--}$&4343& &\\
$3^3P_2$& $2^{++}$&4354& $X(4350)?$&4351(5)&$1^3G_5$& $5^{--}$&4357& &\\
$3^1P_1$& $1^{+-}$&4337& &&$1^1G_4$& $4^{-+}$&4345& &\\
$4^3P_0$& $0^{++}$&4698& &&$1^3H_4$& $4^{++}$&4572& &\\
$4^3P_1$& $1^{++}$&4728& &&$1^3H_5$& $5^{++}$&4592& &\\
$4^3P_2$& $2^{++}$&4763& &&$1^3H_6$& $6^{++}$&4608& &\\
$4^1P_1$& $1^{+-}$&4744& &&$1^3H_5$& $5^{+-}$&4594& &\\
$1^3D_1$& $1^{--}$&3783& $\psi(3770)$&3772.92(35)&\\
$1^3D_2$& $2^{--}$&3795& &&\\
$1^3D_3$& $3^{--}$&3813& &&\\
$1^1D_2$& $2^{-+}$&3807& &&\\
\end{tabular}
\end{ruledtabular}
\end{table}

\begin{table}
\caption{Bottomonium mass spectrum
   (in MeV).} 
   \label{tab:bsm}
\begin{ruledtabular}
\begin{tabular}{cccccccc}
\multicolumn{2}{l}{\underline{\phantom{p}\hspace{1cm}State\hspace{1cm}}}&Theory
&\multicolumn{2}{l}{\underline{\hspace{1.1cm}Experiment \cite{pdg}\hspace{1.1cm}}}&\multicolumn{2}{l}{\underline{\phantom{p}\hspace{1cm}State\hspace{1cm}}}&
Theory\\
$n^{2S+1}L_J$&$J^{PC}$& & meson &mass&$n^{2S+1}L_J$&$J^{PC}$& \\[2pt]
\hline
$1^1S_0$& $0^{-+}$&9398& $\eta_b(1S)$&9390.9(2.8)&$2^3D_1$& $1^{--}$&10435 \\
$1^3S_1$& $1^{--}$&9460& $\Upsilon(1S)$&9460.30(26)&$2^3D_2$& $2^{--}$&10443\\
$2^1S_0$& $0^{-+}$&9990& &&$2^3D_3$& $3^{--}$&10449\\
$2^3S_1$& $1^{--}$&10023& $\Upsilon(2S)$&10023.26(31)&$2^1D_2$& $2^{-+}$&10445\\
$3^1S_0$& $0^{-+}$&10329& &&$3^3D_1$& $1^{--}$&10704\\
$3^3S_1$& $1^{--}$&10355& $\Upsilon(3S)$&10355.2(5)&$3^3D_2$& $2^{--}$&10711\\
$4^1S_0$& $0^{-+}$&10573& &&$3^3D_3$& $3^{--}$&10717\\
$4^3S_1$& $1^{--}$&10586& $\Upsilon(4S)$&10579.4(1.2)&$3^1D_2$& $2^{-+}$&10713\\
$5^1S_0$& $0^{-+}$&10851& &&$4^3D_1$& $1^{--}$&10949\\
$5^3S_1$& $1^{--}$&10869&$\Upsilon(10860)$ &10876(1)&$4^3D_2$& $2^{--}$&10957\\
$6^1S_0$& $0^{-+}$&11061& &&$4^3D_3$& $3^{--}$&10963\\
$6^3S_1$& $1^{--}$&11088&$\Upsilon(11020)$ &11019(8)&$4^1D_2$& $2^{-+}$&10959\\
$1^3P_0$& $0^{++}$&9859& $\chi_{b0}(1P)$&9859.44(52)&$1^3F_2$& $2^{++}$&10343\\
$1^3P_1$& $1^{++}$&9892& $\chi_{b1}(1P)$&9892.78(40)&$1^3F_3$& $3^{++}$&10346\\
$1^3P_2$& $2^{++}$&9912& $\chi_{b2}(1P)$&9912.21(40)&$1^3F_4$& $4^{++}$&10349\\
$1^1P_1$& $1^{+-}$&9900& $h_{b}(1P)$&9898.25(1.50)&$1^1F_3$& $3^{+-}$&10347\\
$2^3P_0$& $0^{++}$&10233&$\chi_{b0}(2P)$ &10232.5(6)&$2^3F_2$& $2^{++}$&10610\\
$2^3P_1$& $1^{++}$&10255&$\chi_{b1}(2P)$ &10255.46(55)&$2^3F_3$& $3^{++}$&10614\\
$2^3P_2$& $2^{++}$&10268& $\chi_{b2}(2P)$&10268.65(55)&$2^3F_4$& $4^{++}$&10617\\
$2^1P_1$& $1^{+-}$&10260&$h_{b}(2P)$ &10259.76(1.57)&$2^1F_3$& $3^{+-}$&10615\\
$3^3P_0$& $0^{++}$&10521& &&$1^3G_3$& $3^{--}$&10511\\
$3^3P_1$& $1^{++}$&10541& &&$1^3G_4$& $4^{--}$&10512\\
$3^3P_2$& $2^{++}$&10550& &&$1^3G_5$& $5^{--}$&10514\\
$3^1P_1$& $1^{+-}$&10544& &&$1^1G_4$& $4^{-+}$&10513\\
$4^3P_0$& $0^{++}$&10781& &&$1^3H_4$& $4^{++}$&10670\\
$4^3P_1$& $1^{++}$&10802& &&$1^3H_5$& $5^{++}$&10671\\
$4^3P_2$& $2^{++}$&10812& &&$1^3H_6$& $6^{++}$&10672\\
$4^1P_1$& $1^{+-}$&10804& &&$1^3H_5$& $5^{+-}$&10671\\
$1^3D_1$& $1^{--}$&10154& &&\\
$1^3D_2$& $2^{--}$&10161&$\Upsilon(1D)$ &10163.7(1.4)&\\
$1^3D_3$& $3^{--}$&10166& &&\\
$1^1D_2$& $2^{-+}$&10163& &&\\
\end{tabular}
\end{ruledtabular}
\end{table}

\begin{table}
\caption{$B_c$ meson mass spectrum
   (in MeV).} 
   \label{tab:bcms}
\begin{ruledtabular}
\begin{tabular}{cccccccc}
\multicolumn{2}{l}{\underline{\phantom{p}\hspace{1cm}State\hspace{1cm}}}&Theory
&\multicolumn{2}{l}{\underline{\hspace{1.1cm}Experiment \cite{pdg}\hspace{1.1cm}}}&\multicolumn{2}{l}{\underline{\phantom{p}\hspace{1cm}State\hspace{1cm}}}&
Theory\\
$n^{2S+1}L_J$&$J^{P}$& & meson &mass&$n^{2S+1}L_J$&$J^{P}$& \\[2pt]
\hline
$1^1S_0$& $0^{-}$&6272& $B_c$&6277(6)&$1^3D_1$& $1^{-}$&7021 \\
$1^3S_1$& $1^{-}$&6333& &&$1D_2$ & $2^{-}$&7025\\
$2^1S_0$& $0^{-}$&6842& &&$1D_2$ & $2^{-}$&7026\\
$2^3S_1$& $1^{-}$&6882& &&$1^3D_3$& $3^{-}$&7029\\
$3^1S_0$& $0^{-}$&7226& &&$2^3D_1$& $1^{-}$&7392\\
$3^3S_1$& $1^{-}$&7258& &&$2D_2$  & $2^{-}$&7399\\
$4^1S_0$& $0^{-}$&7585& &&$2D_2$  & $2^{-}$&7400\\
$4^3S_1$& $1^{-}$&7609& &&$2^3D_3$& $3^{-}$&7405\\
$5^1S_0$& $0^{-}$&7928& &&$3^3D_1$& $1^{-}$&7732\\
$5^3S_1$& $1^{-}$&7947& &&$3D_2$  & $2^{-}$&7741\\
$1^3P_0$& $0^{+}$&6699& &&$3D_2$  & $2^{-}$&7743\\
$1P_1$& $1^{+}$  &6743& &&$3^3D_3$& $3^{-}$&7750\\
$1P_1$& $1^{+}$  &6750& &&$1^3F_2$& $2^{+}$&7273\\
$1^3P_2$& $2^{+}$&6761& &&$1F_3$  & $3^{+}$&7269\\
$2^3P_0$& $0^{+}$&7094& &&$1F_3$  & $3^{+}$&7268\\
$2P_1$  & $1^{+}$&7134& &&$1^3F_4$& $4^{+}$&7277\\
$2P_1$  & $1^{+}$&7147& &&$2^3F_2$& $2^{+}$&7618\\
$2^3P_2$& $2^{+}$&7157& &&$2F_3$  & $3^{+}$&7616\\
$3^3P_0$& $0^{+}$&7474& &&$2F_3$  & $3^{+}$&7615\\
$3P_1$  & $1^{+}$&7500& &&$2^3F_4$& $4^{+-}$&7617\\
$3P_1$  & $1^{+}$&7510& &&$1^3G_3$& $3^{-}$&7497\\
$3^3P_2$& $2^{+}$&7524& &&$1G_4$  & $4^{-}$&7489\\
$4^3P_0$& $0^{+}$&7817& &&$1G_4$  & $4^{-}$&7487\\
$4P_1$  & $1^{+}$&7844& &&$1^3G_5$& $5^{-}$&7482\\
$4P_1$  & $1^{+}$&7853& &&$2G_4$  & $4^{-}$&7819\\
$4^3P_2$& $2^{+}$&7867& &&$2^3G_5$& $5^{-}$&7817\\
\end{tabular}
\end{ruledtabular}
\end{table}

The $B_c$ meson states  with $J=L$, given in
Tables~\ref{tab:bcms},
are  mixtures of spin-triplet $|^3L_L\rangle$  and spin-singlet $|^1L_L\rangle$
states:
\begin{eqnarray}
  \label{eq:mix}
  |\Psi_J\rangle&=&|^1L_L\rangle\cos\theta+|^3L_L\rangle\sin\theta, \cr
 |\Psi'_J\rangle&=&-|^1L_L\rangle\sin\theta+|^3L_L\rangle\cos\theta, \qquad J=L=1,2,3\dots
\end{eqnarray}
where $\theta$ is a mixing angle, and the primed state has the heavier mass.
  Such mixing occurs due to the nondiagonal spin-orbit and
tensor terms in Eq.~(\ref{eq:vsd}). The masses of  physical states were obtained
by diagonalizing the mixing matrix. The obtained values of the mixing
angles $\theta$ are close to the ones given in Ref.~\cite{efg}.

\subsection{Regge trajectories}
\label{sec:rgt}

In our analysis we calculated masses of both orbitally and radially excited
heavy quarkonia up to rather high excitation numbers ($L=5$ and
$n_r=5$). This makes it possible 
to construct the Regge trajectories in the
$(J,M^2)$ and $(n_r,M^2)$ planes using the following  definitions:\\ 
a) $(J,M^2)$ Regge trajectory:
\begin{equation}
  \label{eq:reggej}
J=\alpha M^2+\alpha_0;
\end{equation}

\noindent b)  $(n_r,M^2)$ Regge trajectory:
\begin{equation}
  \label{eq:reggen}
n_r=\beta M^2+\beta_0,
\end{equation}
where $\alpha$, $\beta$ are the slopes and  $\alpha_0$, $\beta_0$ are
the intercepts. The relations (\ref{eq:reggej}) and (\ref{eq:reggen})
arise in most models of quark confinement, but with different values
of the slopes.

\begin{figure}
 \includegraphics[width=13cm]{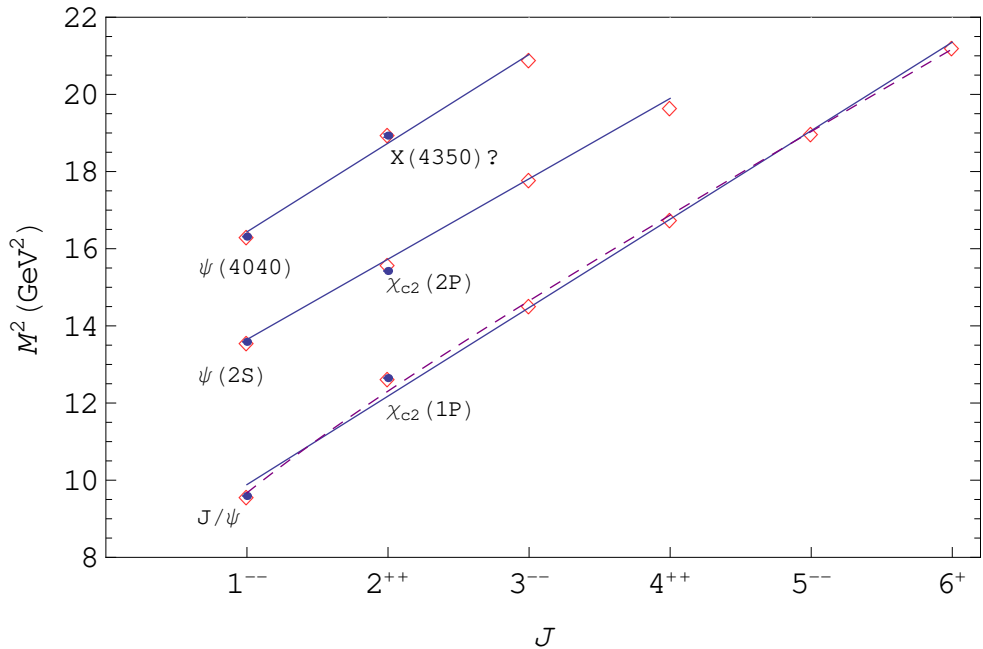} 
\caption{\label{fig:jpsi} Parent and daughter ($J, M^2$) Regge trajectories for
  charmonium states  with natural parity ($P=(-1)^J$). Diamonds are predicted
  masses. Available experimental data are given by dots with  particle
  names. The dashed line  corresponds to a nonlinear fit for the parent trajectory.} 
\end{figure}

\begin{figure}
  \includegraphics[width=13cm]{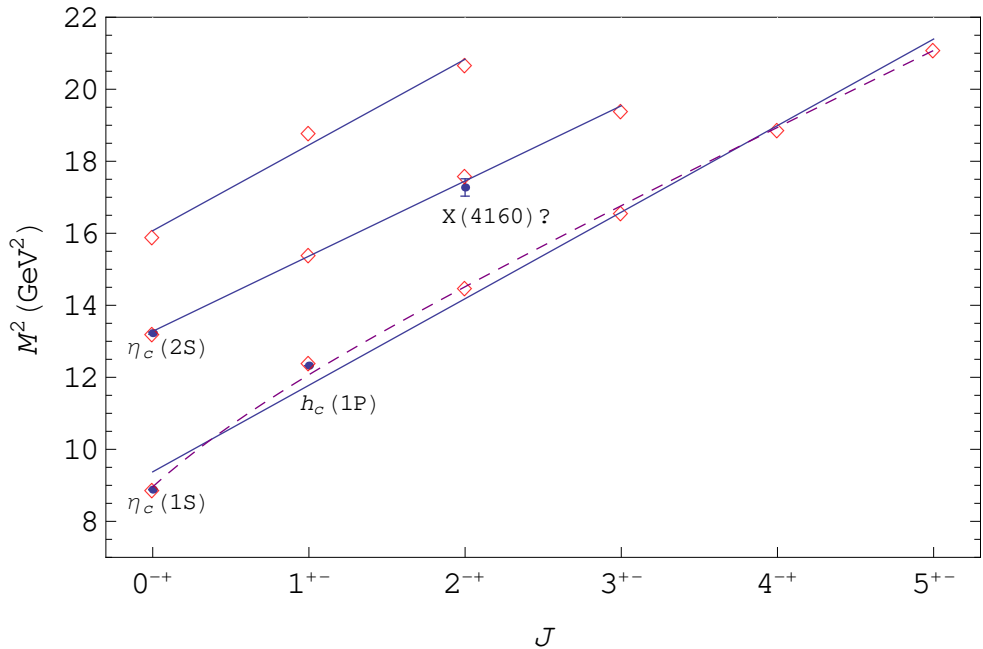}

\caption{\label{fig:etac} Same as in Fig.~\ref{fig:jpsi} for
  charmonium states with unnatural parity ($P=(-1)^{J+1}$). }
\end{figure}

\begin{figure}
  \includegraphics[width=13cm]{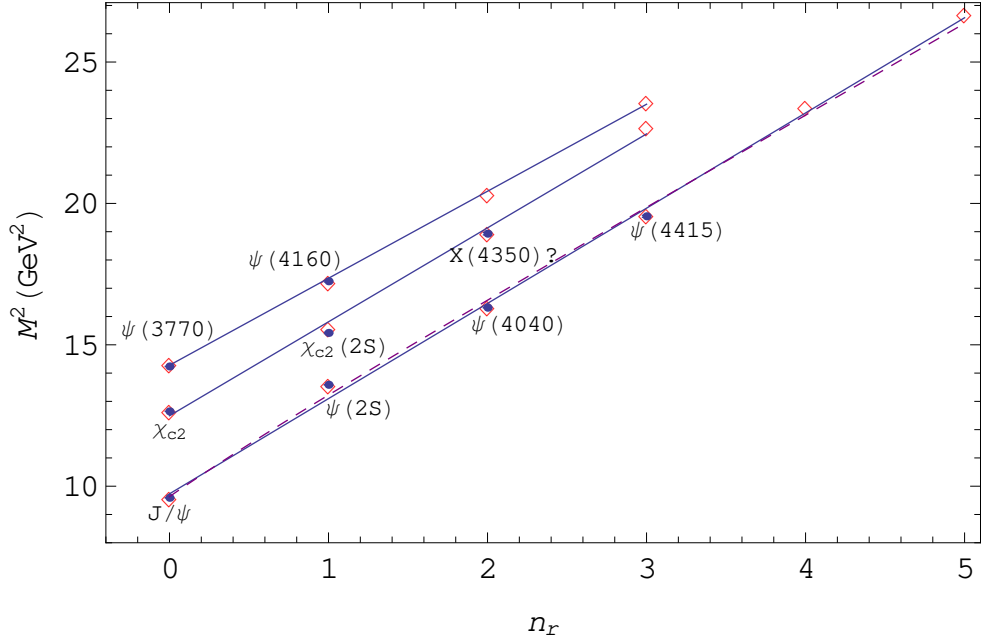}

\caption{\label{fig:psin} The $(n_r,M^2)$ Regge trajectories for
  vector ($S$-wave), tensor and vector ($D$-wave) charmonium states (from bottom to
  top). Notations are the same as in Fig.~\ref{fig:jpsi}. }
\end{figure}

\begin{figure}

\includegraphics[width=13cm]{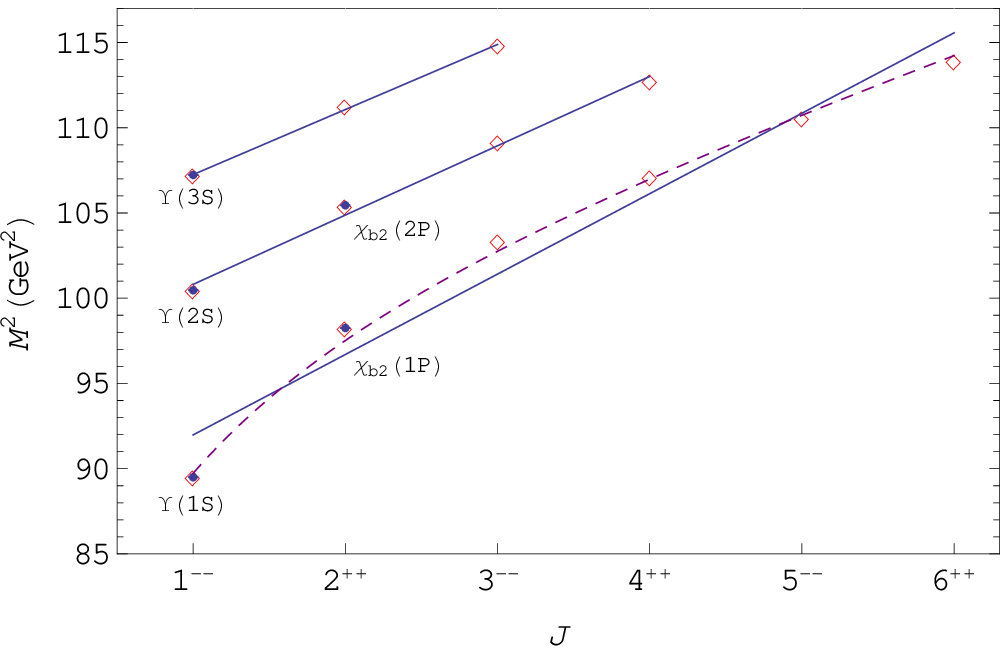} 
\caption{\label{fig:upsilon} Same as in Fig.~\ref{fig:jpsi} for
  bottomonium states with natural parity.  }
\end{figure}

\begin{figure}
  \includegraphics[width=13cm]{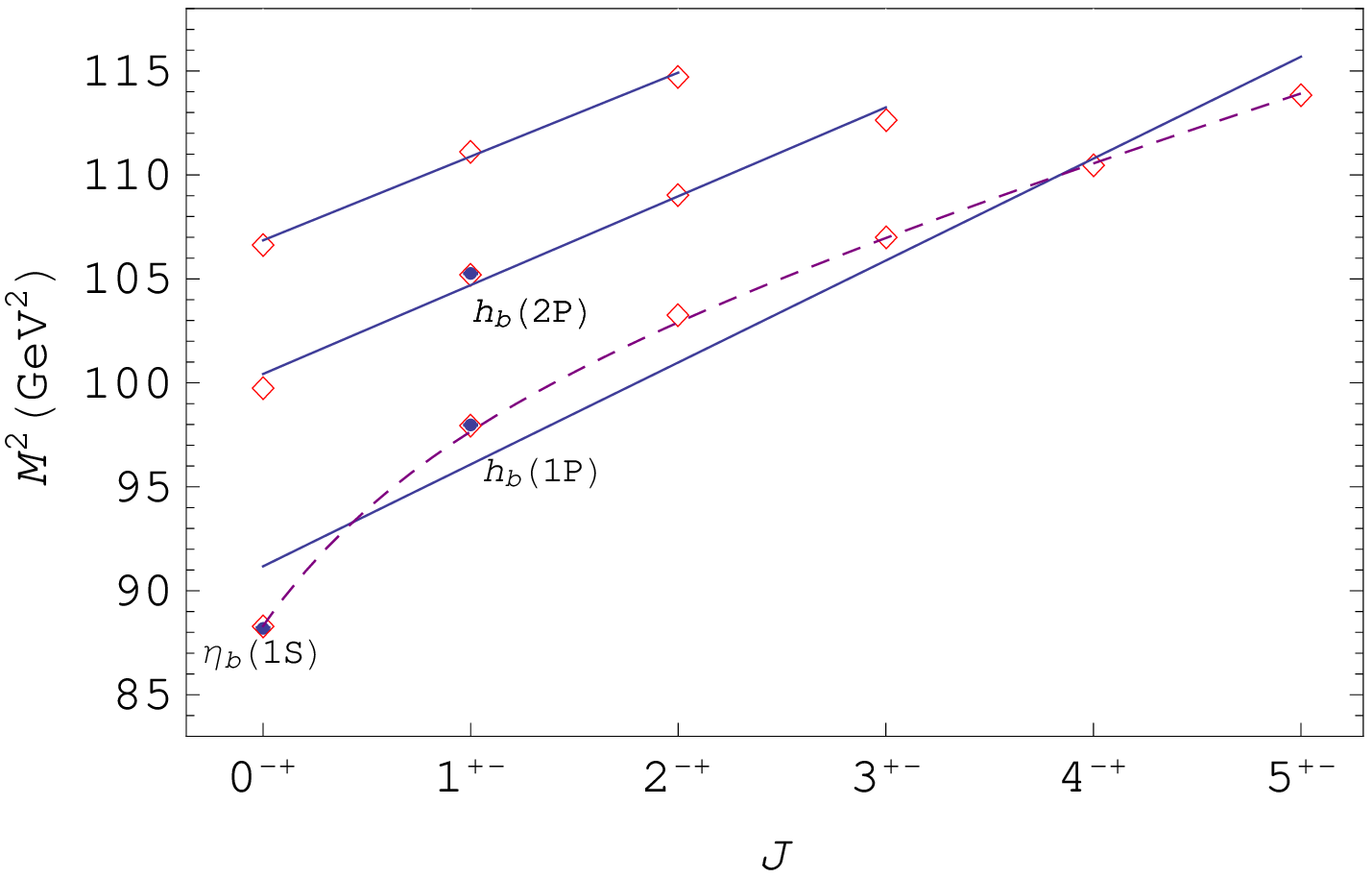}

\caption{\label{fig:etab} Same as in Fig.~\ref{fig:jpsi} for
  bottomonium states with unnatural parity. }
\end{figure}

\begin{figure}
 \includegraphics[width=13cm]{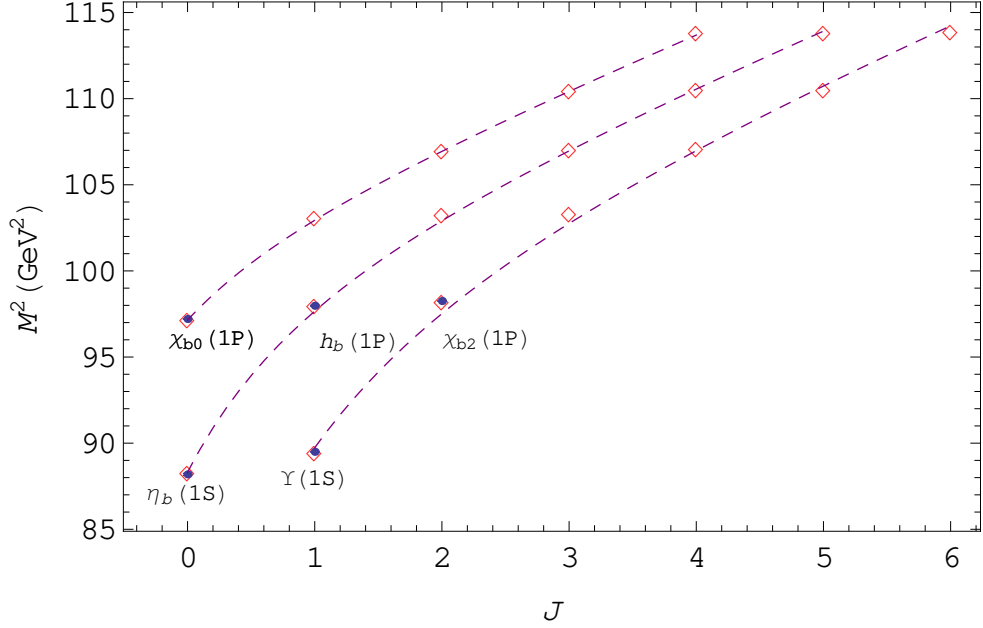} 
\caption{\label{fig:upsnl}  Nonlinear Regge trajectories in the ($J,
  M^2$) starting from
  vector, pseudoscalar and scalar bottomonium states (from bottom to
  top). } 
\end{figure}

\begin{figure}
  \includegraphics[width=13cm]{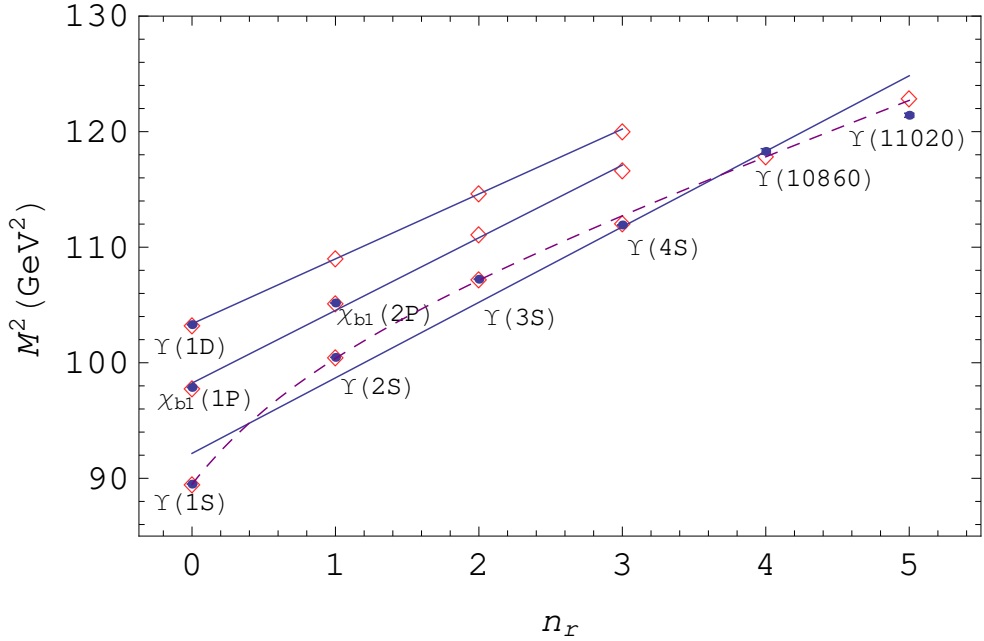}

\caption{\label{fig:upsilonn} Same as in Fig.~\ref{fig:psin} for
  bottomonium.  }
\end{figure}

\begin{figure}
 \includegraphics[width=13cm]{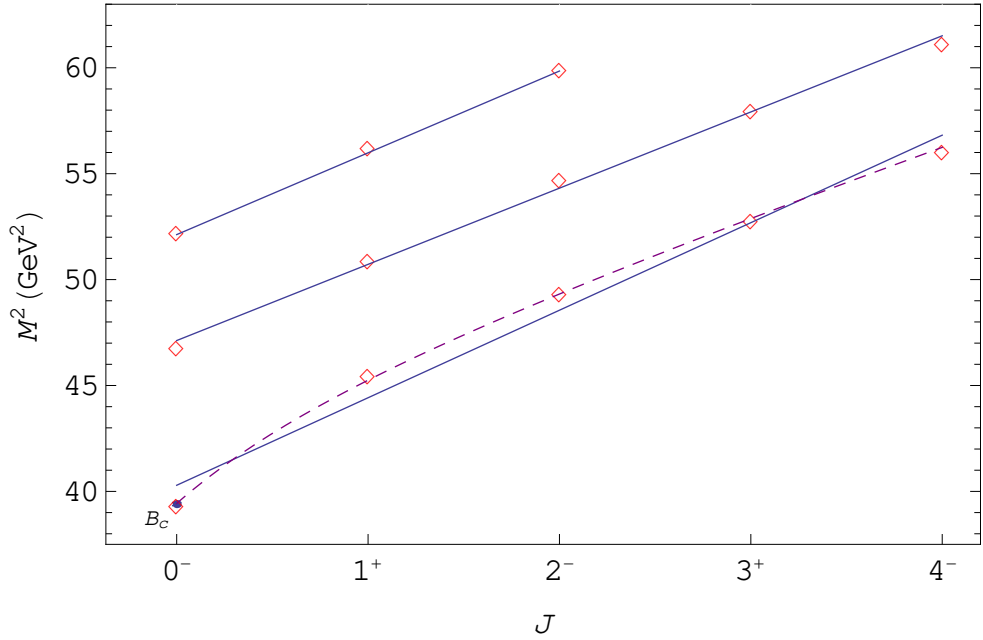}

\caption{\label{fig:Bc} Same as in Fig.~\ref{fig:jpsi} for
  $B_c$ meson states  with unnatural parity. }
\end{figure}

\begin{figure}
  \includegraphics[width=13cm]{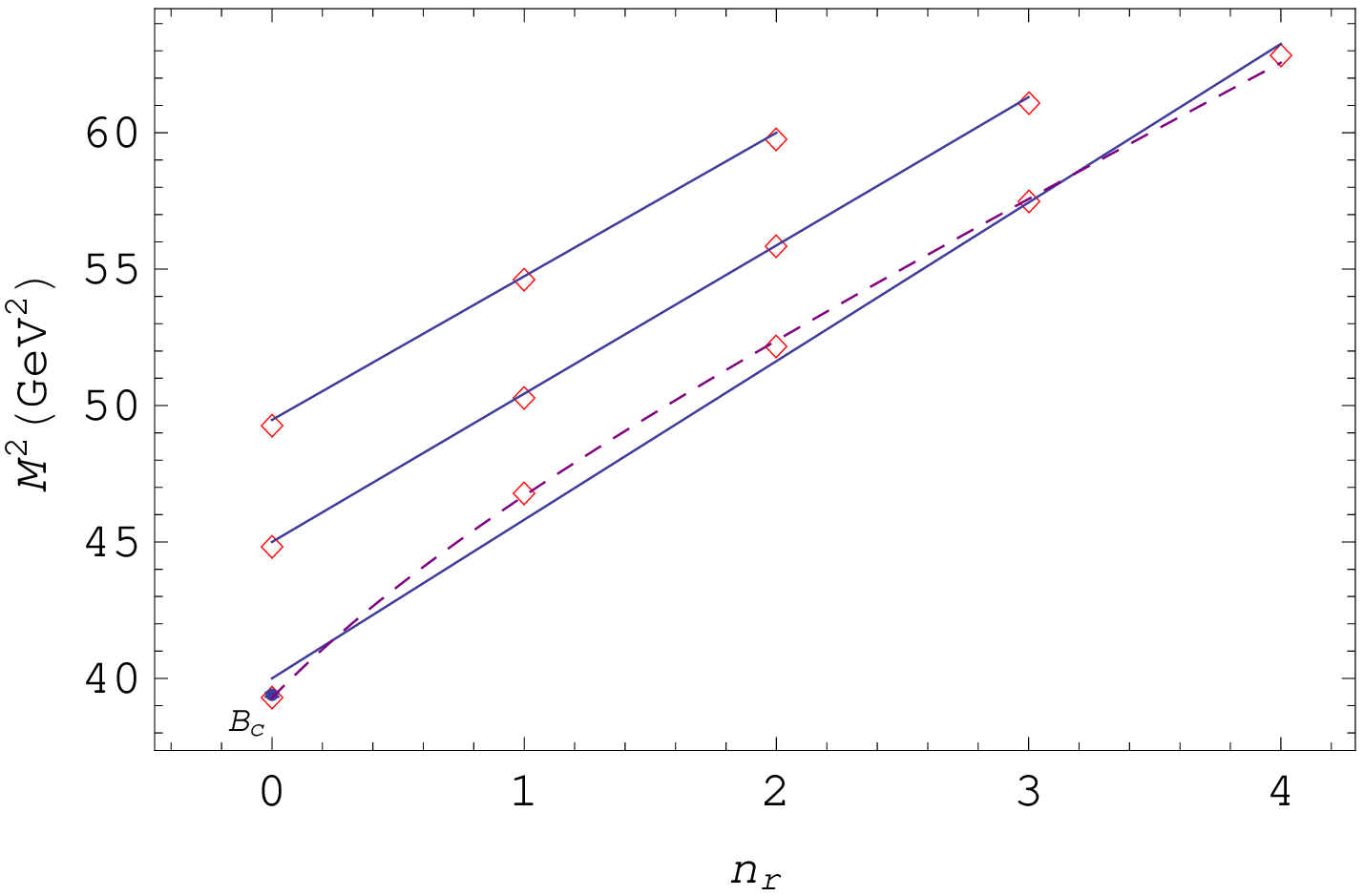}

\caption{\label{fig:Bcn} Same as in Fig.~\ref{fig:psin} for
 the  $B_c$ meson.  }
\end{figure}

\begin{figure}
  \includegraphics[width=13cm]{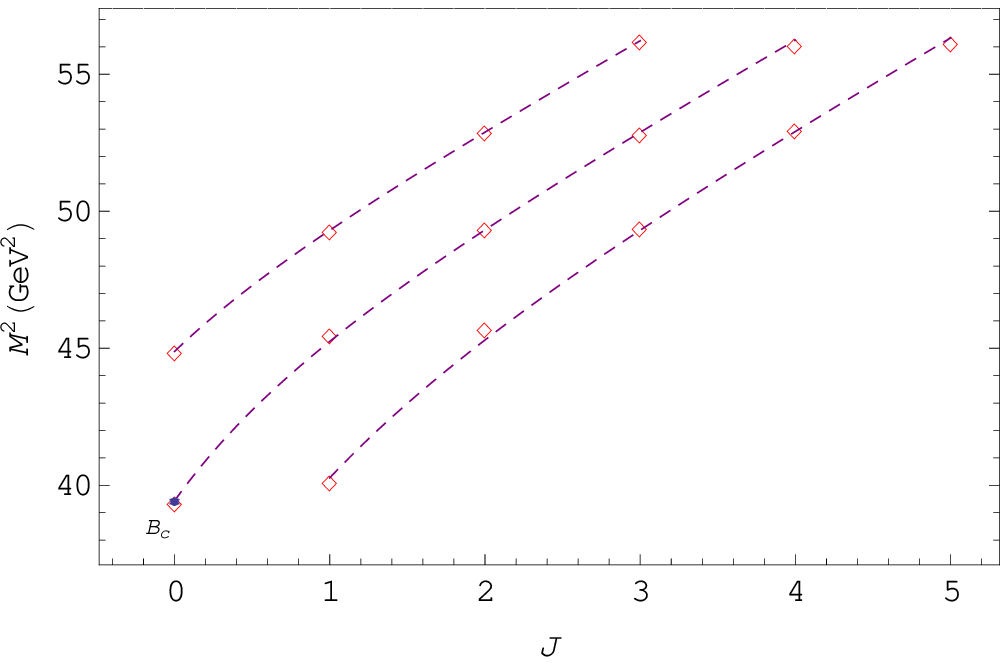}

\caption{\label{fig:bcnlj} Same as in Fig.~\ref{fig:upsnl} for the  $B_c$ meson. }
\end{figure}

\begin{table}
  \caption{Fitted parameters of the $(J,M^2)$ parent and daughter Regge
    trajectories for heavy quarkonia and $B_c$ mesons 
    with natural and unnatural parity.} 
  \label{tab:rtj}
\begin{ruledtabular}
\begin{tabular}{ccccc}
Trajectory& 
\multicolumn{2}{l}{\underline{\hspace{1.6cm}natural
    parity\hspace{1.6cm}}}&
 \multicolumn{2}{l}{\underline{\hspace{1.6cm}unnatural
    parity\hspace{1.6cm}}}\hspace{-.1cm}\\ 
&$\alpha$ (GeV$^{-2}$)& $\alpha_0$&$\alpha$  (GeV$^{-2}$)&$\alpha_0$\\
\hline
$c\bar c$ &$J/\psi$&&$\eta_c$\\
parent& $0.436\pm0.014$& $-3.31\pm0.22$& $0.416\pm0.021$&
$-3.90\pm0.31$\\
first daughter &$0.488\pm0.011$&$-5.63\pm0.18$ & $0.479\pm0.015$&
$-6.36\pm0.24$\\
second daughter &$0.431\pm0.036$&$-6.08\pm0.68$ & $0.414\pm0.050$&
$-6.66\pm0.92$\\
$c\bar c$ &$\chi_{c0}$&&$\chi_{c1}$\\
parent& $0.431\pm0.016$& $-5.07\pm0.25$& $0.461\pm0.008$&
$-4.66\pm0.12$\\
daughter &$0.493\pm0.031$&$-7.41\pm0.53$ & $0.456\pm0.006$&
$-5.83\pm0.11$\\
$b\bar b$ &$\Upsilon$&&$\eta_b$\\
parent& $0.212\pm0.022$& $-18.5\pm2.3$& $0.184\pm0.024$&
$-16.7\pm2.5$\\
first daughter &$0.246\pm0.014$&$-23.8\pm1.5$ & $0.234\pm0.016$&
$-23.5\pm1.7$\\
second daughter &$0.262\pm0.010$&$-27.1\pm1.1$ & $0.248\pm0.014$&
$-26.5\pm1.6$\\
third daughter &$0.246\pm0.027$&$-26.6\pm3.1$ & $0.241\pm0.026$&
$-27.0\pm3.0$\\
$b\bar b$ &$\chi_{b0}$&&$\chi_{b1}$\\
parent& $0.228\pm0.021$& $-22.3\pm2.2$& $0.239\pm0.018$&
$-22.5\pm1.9$\\
daughter &$0.254\pm0.009$&$-26.7\pm1.0$ & $0.267\pm0.006$&
$-27.1\pm0.7$\\
$b\bar c$ &$B^*_c$&&$B_c$\\
parent& $0.254\pm0.018$& $-9.38\pm0.88$& $0.242\pm0.019$&
$-9.75\pm0.95$\\
first daughter &$0.291\pm0.008$&$-12.8\pm0.4$ & $0.278\pm0.009$&
$-13.1\pm0.5$\\
second daughter &$0.270\pm0.010$&$-13.3\pm0.4$ & $0.259\pm0.007$&
$-13.5\pm0.4$\\
$b\bar c$ &$B_{c0}$&&$B_{c1}$\\
parent& $0.265\pm0.013$& $-12.0\pm0.6$& $0.285\pm0.007$&
$-13.0\pm0.4$\\
daughter &$0.275\pm0.014$&$-13.9\pm0.8$ & $0.298\pm0.008$&
$-15.3\pm0.4$\\ 
\end{tabular}
 \end{ruledtabular}
\end{table}

\begin{table}
  \caption{Fitted parameters of the $(n_r,M^2)$  Regge
    trajectories for heavy quarkonia and $B_c$ mesons.} 
  \label{tab:rtn}
\begin{ruledtabular}
\begin{tabular}{cccccc}
Meson&$\beta$ (GeV$^{-2}$)& $\beta_0$&Meson&$\beta$ (GeV$^{-2}$)& $\beta_0$\\
\hline
$c\bar c$&&& $c\bar c$\\
$\eta_c$& $0.287\pm0.011$& $-2.62\pm0.18$&$J/\psi$& $0.297\pm0.010$&$-2.89\pm0.16$\\
$\chi_{c0}$&$0.288\pm0.003$&$-3.34\pm0.06$ &$\chi_{c1}$& $0.301\pm0.011$&$-3.67\pm0.19$\\
$\chi_{c2}$& $0.301\pm0.011$&$-3.76\pm0.019$ &$h_c$& $0.298\pm0.010$&$-3.68\pm0.17$\\
$\psi(^3D_1)$& $0.325\pm0.006$&$-4.62\pm0.11$&$\psi(^3D_2)$&$0.315\pm0.003$& $-4.53\pm0.06$\\
$\psi(^3D_3)$& $0.311\pm0.002$&$-4.53\pm0.04$&$\psi(^1D_2)$& $0.317\pm0.003$&$-4.53\pm0.06$\\

$b\bar b$&&& $b\bar b$\\
$\eta_b$& $0.151\pm0.013$& $-13.7\pm1.4$&$\Upsilon$& $0.153\pm0.012$&$-14.1\pm1.3$\\
$\chi_{b0}$&$0.158\pm0.008$&$-15.4\pm0.9$ &$\chi_{b1}$& $0.159\pm0.007$&$-15.7\pm0.8$\\
$\chi_{b2}$& $0.161\pm0.007$&$-15.9\pm0.7$ &$h_b$& $0.161\pm0.007$&$-15.8\pm0.8$\\
$\Upsilon(^3D_1)$& $0.178\pm0.002$&$-18.4\pm0.3$&$\Upsilon(^3D_2)$&$0.178\pm0.002$& $-18.4\pm0.3$\\
$\Upsilon(^3D_3)$& $0.178\pm0.003$&$-18.4\pm0.3$&$\Upsilon(^1D_2)$& $0.178\pm0.002$&$-18.4\pm0.3$\\
$b\bar c$&&& $b\bar c$\\
$B_c$& $0.172\pm0.008$& $-6.88\pm0.39$&$B_c^*$& $0.175\pm0.008$&$-7.15\pm0.39$\\
$B_{c0}$&$0.184\pm0.001$&$-8.28\pm0.07$ &$B_{c2}$& $0.185\pm0.001$&$-8.48\pm0.07$\\
$B_c(^3D_1)$& $0.190\pm0.002$&$-9.40\pm0.12$&$B_c(^3D_2)$&$0.188\pm0.002$& $-9.27\pm0.11$\\
\end{tabular}
 \end{ruledtabular}
\end{table}

\begin{figure}
  \includegraphics[width=13cm]{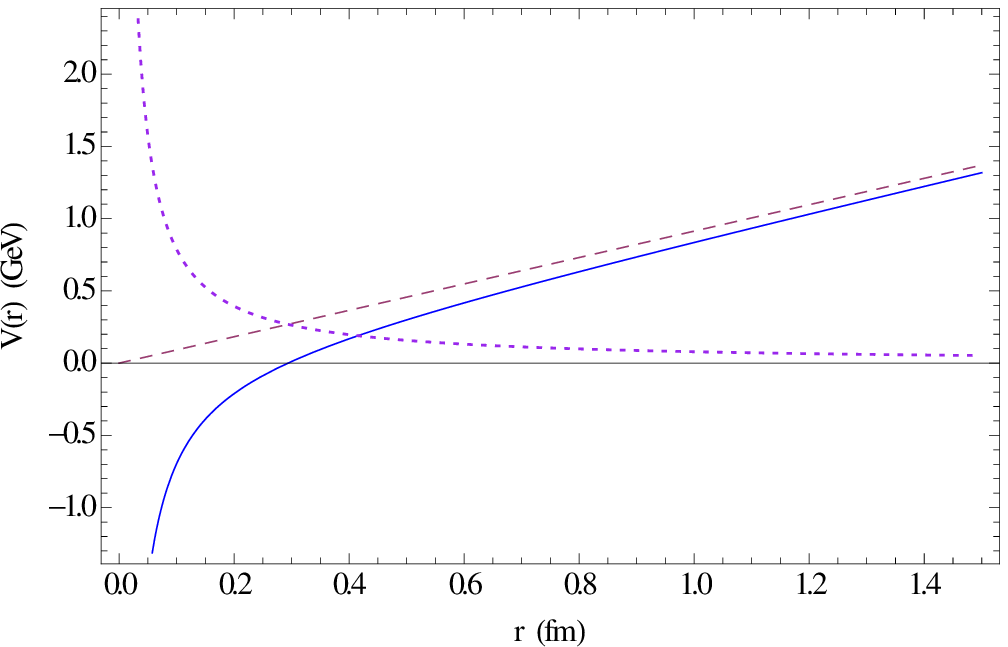}

\caption{\label{fig:pot} Static potential of the quark-antiquark
  interaction (\ref{cp}) without the constant term (solid line). Dashed line shows
  the linear confining potential contribution, while dotted line corresponds to
  the modulus of the Coulomb potential. }
\end{figure}

In Figs.~\ref{fig:jpsi}-\ref{fig:bcnlj} we plot the Regge trajectories in
the ($J, M^2$) and $(n_r,M^2)$ planes for  charmonia, bottomonia and
$B_c$ mesons.  The masses calculated in our
model are shown by diamonds. Available experimental data are given by
dots with error bars and corresponding meson names. 
Straight lines were obtained by the
$\chi^2$ fits of the calculated values. The fitted slopes
and intercepts of the Regge trajectories are given in
Tables~\ref{tab:rtj} and \ref{tab:rtn}. We see that the calculated
charmonium masses fit nicely to the linear trajectories in both
planes  (maybe with the exception of the parent trajectories, where
the  $J/\psi$ and 
$\eta_c$ mesons seem to have slightly lower masses). These
trajectories are almost parallel and equidistant. In the 
bottomonium and $B_c$ meson sectors the situation is more complicated. The
daughter trajectories, which involve both radially and orbitally
excited states, turn out to be almost linear. On the other hand, the parent
trajectories, which start from ground states, are exhibiting a
nonlinear behaviour in the lower mass (excitation) region. Such
nonlinearity is more pronounced in bottomonium. The origin of this
nonlinearity can be easily understood, if one compares the mean radii
of these states. The values of the mean square radii $\sqrt{\langle
  r^2\rangle}$ of charmonia, $B_c$ mesons and bottomonia, calculated in
our model, are given in Table~\ref{tab:msr}. The static potential of
the quark-antiquark interaction is plotted in Fig.~\ref{fig:pot}
(solid line). In
this figure we also separately plot the contributions from linear
confinement (dashed line) and of the modulus of the Coulomb
potential (dotted line). As seen form Fig.~\ref{fig:pot}, the
Coulomb potential dominates for distances less than 0.15~fm, while
the confining potential is dominant for distances larger than
0.5~fm. In the intermediate region both potentials play an equally
important role. Therefore the light mesons and charmonia (with the exception of
the $\eta_c$ and $J/\psi$ which are in the intermediate region) have 
characteristic sizes which belong to the region, where the confining
potential dominates in the interquark  potential
(\ref{cp}). This leads to the emergence of the linear Regge
trajectories. Contrary, the ground and few first excited states of bottomonia
and $B_c$ mesons have smaller sizes and fall into the region, where the
Coulomb part of the potential (\ref{cp}) gives an important
contribution. As a 
result, the parent Regge trajectories of bottomonia and $B_c$ mesons
are nonlinear, while the daughter trajectories (which fall into the region,
where the confining potential is dominant) are still linear ones. In
Refs.~\cite{gll,serg} an interpolating formula between the limiting cases
of pure Coulomb and linear interactions was proposed. It can be
written as follows:\\ 
(a) for the parent trajectory in the ($J, M^2$) plane
\begin{equation}
  \label{eq:nlj}
 M^2=\left(J-\frac{\gamma_1}{(J+2)^2}+\gamma_0\right)/\gamma,
\end{equation}
(b) for the $J=1$ trajectory in the ($ n_r, M^2$) plane
\begin{equation}
  \label{eq:nlj}
  M^2=\left(n_r-\frac{\tau_1}{(n_r+2)^2}+\tau_0\right)/\tau,
\end{equation}
where the parameters $\gamma$, $\tau$, $\gamma_0$, $\tau_0$ and
$\gamma_1$, $\tau_1$ determine the slopes, intercepts and nonlinearity
of the Regge trajectories, respectively. Their fitted values are given in
Table~\ref{tab:nlrt}. The corresponding Regge trajectories are plotted in
Figs.~\ref{fig:jpsi}-\ref{fig:bcnlj} by dashed lines. It is found that
these nonlinear trajectories have the same slope $\gamma$ for the given
quarkonium family, which is generally in agreement (but slightly higher)
with the linear trajectory slopes $\alpha$ for the respective daughter
trajectories given in Table~\ref{tab:rtj}. We see that the
nonlinearity of the charmonium Regge trajectories is almost
negligible, and its account does not significantly improve the quality of the fit
compared to the linear one.

\begin{table}
  \caption{Mean square radii $\sqrt{\langle r^2\rangle}$ for the spin-singlet ground
  and excited states of charmonia, $B_c$ mesons and bottomonia (in fm).}  
  \label{tab:msr}
\begin{ruledtabular}
\begin{tabular}{cccc}
State& $\sqrt{\langle r^2\rangle_\psi}$  & $\sqrt{\langle
  r^2\rangle_{B_c}}$ & $\sqrt{\langle r^2\rangle_\Upsilon}$\\
\hline
$1S$ & 0.37 & 0.33 & 0.22\\
$1P$ & 0.59 & 0.53 & 0.41\\
$2S$ & 0.71 & 0.63 & 0.50\\
$1D$ & 0.74 & 0.67 & 0.54\\
$2P$ & 0.87 & 0.79 & 0.65\\
$1F$ & 0.87 & 0.79 & 0.65\\
$3S$ & 0.94 & 0.87 & 0.72\\
$1G$ & 0.98 & 0.89 & 0.75\\
$2D$ & 0.99 & 0.90 & 0.76\\
$1H$ & 1.08 & 0.99 & 0.85\\
$3P$ & 1.09 & 0.99 & 0.84\\
$2F$ & 1.09 & 0.99 & 0.85\\
$4S$ & 1.16 & 1.05 & 0.90\\
$3D$ & 1.18 & 1.08 & 0.94\\
$4P$ & 1.26 & 1.16 & 1.01\\
$5S$ & 1.32 & 1.21 & 1.07\\
$6S$ & 1.46 &      & 1.22\\
\end{tabular}
 \end{ruledtabular}
\end{table}

We can compare the slopes of linear Regge trajectories for heavy
quarkonia obtained in this paper with our previous results for the
slopes of Regge trajectories of light \cite{lregge} and heavy-light
\cite{hlrt} mesons.  Such comparison shows that the slopes decrease
rather fast with the growth of the quark masses: the slope $\alpha$
decreases from about 1.1~GeV$^{-2}$ for light 
mesons, composed from $u,d$ quarks and antiquarks, to about
0.24~GeV$^{-2}$ for bottomonium. However, the difference between
slopes of heavy-light ($Q\bar q$) mesons and of heavy quarkonia ($Q\bar Q'$)
is not so dramatic. In fact, comparing the present Tables~\ref{tab:rtj}, 
\ref{tab:rtn} and the corresponding Tables~4, 5 of Ref.~ \cite{hlrt}, we see that the
slopes for charmonia and $D$, $D_s$ mesons  have very close values, and the
same is true also for bottomonia, $B_c$ and $B$, $B_s$  mesons. This
might indicate that the slope of the meson Regge trajectory is mainly determined by
the mass of the heaviest quark $m_Q$. The dependence of the Regge
slopes $\alpha$ and $\beta$ on $m_Q$ has  in both planes with rather good
accuracy the same simple form:
$\alpha, \beta\propto 1/\sqrt{m_Q}$.  
    
\begin{table}
  \caption{Fitted parameters of the nonlinear Regge
    trajectories for heavy quarkonia and $B_c$ mesons.} 
  \label{tab:nlrt}
\begin{ruledtabular}
\begin{tabular}{ccccccc}
Meson&$\gamma$ (GeV$^{-2}$)& $\gamma_0$& $\gamma_1$&$\tau$
(GeV$^{-2}$)& $\tau_0$& $\tau_1$\\
\hline
$\Upsilon$ & 0.33 & 32.2 & 32.3& 0.22 & 22.2 & 10.1\\
$\eta_b$   & 0.33 & 32.9 & 15.0\\
$\chi_{b0}$ & 0.33 & 33.7 & 6.57\\
$B_c^*$    & 0.32 & 13.3 & 12.5 & 0.21 & 9.25 & 4.00\\
$B_c$      & 0.32 & 14.2 & 6.21 \\
$B_{c0}$   & 0.32 & 15.1 & 2.99\\
$J/\psi$ & 0.48 & 4.25 & 5.47& 0.31 & 3.19 & 0.82\\
$\eta_c$   & 0.48 & 5.19 & 3.56\\
\end{tabular}
 \end{ruledtabular}
\end{table}

From the comparison of the slopes in Tables~\ref{tab:rtj},
\ref{tab:rtn} we see that the $\alpha$ values are systematically larger than
the $\beta$ ones. The ratio of their mean values is about 1.4 for
charmonia, bottomonia and $B_c$ mesons. This value of the ratio is the
same as for the charmed and bottom mesons \cite{hlrt}, but slightly
larger than for light mesons \cite{lregge}, for which
$\alpha/\beta$ was found to be about 1.3. 

\subsection{Comparison with experiment}
\label{sec:ce}
In Tables~\ref{tab:csm}-\ref{tab:bcms} we compare our predictions for
the heavy quarkonium masses with the available experimental data
\cite{pdg}. We find  that all states below the open flavour
production thresholds are well described by our model, the difference
between predicted and measured masses does not exceed a few MeV. For
higher excited states, which are above this threshold, most of the
well-established conventional states (believed to be quark-antiquark ones) are also
well described by our model, the 
difference between theory and experiment being somewhat larger, but it still
does not exceed 20 MeV.~\footnote{Note that hadron loop effects
  \cite{eichten,bs} can lead to mass shifts and state mixings. As 
  shown in Ref.~\cite{bs}, the loop mass shifts can be absorbed by a
  change of the valence quark model parameters. However, the opening
  of the new threshold could provide a larger mass shifts to the nearby
quarkonium states.} As it is seen from
Figs.~\ref{fig:jpsi}-\ref{fig:bcnlj}, these states fit to the
corresponding Regge trajectories both in the $(J,M^2)$ and
$(n_r,M^2)$ planes. 

We first discuss the recently found quarkonium states below the open
flavour production threshold.
The observation and measurement of the mass of the pseudoscalar
ground state $\eta_b$ \cite{etab} provides a significant information
about the spin-spin interaction in heavy quarkonia. The averaged
bottomonium hyperfine splitting measured in
$\Upsilon(3S)\to\eta_b(1S)\gamma$ and
$\Upsilon(2S)\to\eta_b(1S)\gamma$ decays  is $\Delta M_{\rm hfs}(\eta_b)\equiv
M_{\Upsilon(1S)}-M_{\eta_b(1S)}= 69.3 \pm 2.8$~MeV
\cite{pdg,etab}. Very recently the Belle Collaboration \cite{belleetab} reported the
first observation of 
the radiative transition $h_b(1P)\to\eta_b(1S)\gamma$. The measured
$\eta_b(1S)$ mass is $9401.0\pm1.9^{+1.4}_{-2.4}$~MeV and the hyperfine
splitting $\Delta M_{\rm hfs}(\eta_b)=59.3\pm1.9^{+2.4}_{-1.4}$~MeV
\cite{belleetab}. Our prediction for 
this splitting, $\Delta M_{\rm hfs}(\eta_b)=62$~MeV, is in good agreement
with the experimental values.~\footnote{Almost the same value of
  $\Delta M_{\rm hfs}(\eta_b)=60$~MeV was predicted by us in Ref.~\cite{efg},
  while most of other theoretical predictions, especially the ones
  based on perturbative calculations, gave significantly lower central
  values, e.g.,  $41\pm14$~MeV \cite{pertetab}.}  Note that our model
correctly predicts the branching ratios of the corresponding radiative
decays \cite{efg}.
For  the better understanding of the hyperfine interaction in heavy
mesons it will be very interesting 
to measure the mass of the vector  $B_c^*$-meson, which consists
of two heavy quarks of different flavours, and determine
the related hyperfine splitting. Our model correctly predicts the pseudoscalar
$B_c$ meson mass and gives for the hyperfine splitting the value of
$\Delta M_{\rm hf}(B_c)=61$~MeV. Note that the LHCb Collaboration very
recently measured the $B_c$ meson mass, and their preliminary result
is $M(B^+_c) = 6268.0\pm4.0({\rm stat})\pm0.6({\rm syst})$~MeV,
leading to the improved average $M(B^+_c)^{{\rm exp}} =6272.95\pm
5.17$~MeV \cite{lhcb}.

Another important experimental test of the structure of the spin
splittings in heavy quarkonia comes from the measurement of the
masses of the spin-singlet $P$-levels first in charmonium $h_c(1P)$
\cite{exphc} and very recently in bottomonium $h_b(1P)$ 
and $h_b(2P)$ \cite{exphb}. The measured masses of these states
almost coincide with the spin-averaged centroid of the triplet states
$\langle
M(^3P_J)\rangle=[M(\chi_{Q0})+3M(\chi_{Q1})+5M(\chi_{Q2})]/9$. The
hyperfine mass splittings $\Delta M_{\rm hfs}(nP)\equiv \langle
M(n^3P_J)\rangle-M(n^1P_1)$ in bottomonium are found to be $\Delta M_{\rm hfs}(1P)=(1.62\pm
1.52)$~MeV and $\Delta M_{\rm hfs}(2P)=(0.48^{+1.57}_{-1.22})$~MeV  \cite{exphb}.  This
observation indicates that the spin-spin contribution is negligible for
$P$-levels, and thus shows the vanishing of the long-range chromomagnetic 
interaction in heavy quarkonia. In our model this is the result of the
choice of the value of the long-range chromomagnetic quark moment
$\kappa=-1$. Note that our original predictions \cite{efg} for the
spin-singlet masses are confirmed by these measurements.    

The recently observed $\Upsilon(1^3D_2)$ state is the only $D$-wave state
found below the threshold of open flavour production. Our prediction
for its mass is consistent with the measured value. It will be
interesting to observe other $\Upsilon(1D)$ states in order to test
further our understanding of spin-orbit and spin-spin interactions in
heavy quarkonia.

Next we discuss the observed states above the open flavour production
threshold. The most well-established states are the vector $1^{--}$
states. For charmonium PDG \cite{pdg} lists seven such states: $\psi(3770)$,
$\psi(4040)$, $\psi(4160)$, $X(4260)$, $X(4360)$, $\psi(4415)$ and
$X(4660)$, from which only the $\psi$ states are included in the PDG Summary
Tables \cite{pdg}. These states are believed to be ordinary $c\bar c$ charmonium
(with isospin $I=0$). They are well described by our
model (see Table~\ref{tab:csm}): $\psi(4040)$ and $\psi(4415)$ are the
$3^3S_1$ and $4^3S_1$ states, while  $\psi(3770)$ and  $\psi(4160)$
are the $1^3D_1$ and $2^3D_1$ states, respectively. These $\psi$ states fit
well to the corresponding Regge trajectories (see
Fig.~\ref{fig:psin}). On the other hand, the
three new vector states $X$ are considered as unexpected
exotic states (their isospin is not determined
experimentally). Indeed, we do not have any $c\bar c$ candidates for these
states in Table~\ref{tab:csm}. Contrary, in Ref.~\cite{htetr} we have found that
these states  can be described in our model as tetraquarks
composed from a diquark and antidiquark ($[cq][\bar c \bar q]$,
$q=u,d$). In particular, the  $X(4260)$ and
$X(4660)$ states can be interpreted as the $1^{--}$
states of such tetraquarks  with a scalar diquark $[cq]_{S=0}$ and scalar antidiquark
$[\bar c \bar q]_{S=0}$ in the relative  $1P$- and $2P$-states and
predicted masses 4244~MeV and 4666~MeV,
respectively \cite{htetr}. The  $X(4360)$ can be viewed as the $1^{--}$
tetraquark with the axial vector diquark $[cq]_{S=1}$ and axial vector antidiquark
$[\bar c \bar q]_{S=1}$  in the relative $1P$-state, which mass is
predicted to be 4350~MeV \cite{htetr}.~\footnote{Note that in
  Ref.~\cite{vento} a different prescription for these states is
  used: $X(4260)$, $X(4360)$, $\psi(4415)$ and $X(4660)$ are
  assigned to $4S$, $3D$, $5S$ and $6S$ charmonium $c\bar c$
  states. This leads the authors to the conclusion that the quark
  interaction potential should be screened at large distances.}   

The three vector bottomonium states, $\Upsilon(10580)$,
$\Upsilon(10860)$ and $\Upsilon(11020)$,  observed above open bottom threshold
\cite{pdg},
are rather well described in our model as $4^3S_1$, $5^3S_1$ and $6^3S_1$ states (see
Table~\ref{tab:bsm}). The mass of $\Upsilon(11020)$ being somewhat
higher than the experimental value. They fit to the corresponding Regge trajectory
in Fig.~\ref{fig:upsilonn}.  

The only experimentally established $2P$ charmonium state is
$\chi_{c2}(2P)$ which mass is predicted slightly higher (by about 
20~MeV) in our model. From Table~\ref{tab:csm} we see that the exotic
 state $X(3872)$ cannot be described as the $1^{++}$ $2^3P_1$ $c\bar c$
state or the $2^{-+}$ $1^1D_2$ $c\bar c$ state. If this state
belonged to either $2P$ or $1D$ multiplets, this could signal a large fine
splitting in these multiplets, since the $X(3872)$ mass is 55~MeV below
$\chi_{c2}(2P)$ and 100~MeV above $\psi(3770)$. As we see from
Table~\ref{tab:csm}, our model does not support such large fine splittings.  In
Ref.~\cite{tetr,htetr} we argued that $X(3872)$ can be considered as the
$1^{++}$ ground state tetraquark, composed from the scalar and axial
vector diquark and antidiquark ($([cq]_{S=0}[\bar c \bar
q]_{S=1}+[cq]_{S=1}[\bar c \bar q]_{S=0})/\sqrt2)$), which mass is
predicted to be 3871~MeV.  

As we see from Table~\ref{tab:csm}, the $X(4160)$ and $X(4350)$ can be
attributed from the point of view of the mass value and charge parity $C=+$
to the pseudo tensor $2^{-+}$ spin-singlet $2^1D_2$  
and tensor $2^{++}$ spin-triplet $3^3P_2$ charmonium states,
respectively. They fit well to the corresponding Regge trajectories in
Figs.~\ref{fig:jpsi}-\ref{fig:psin}. 

The $X(4140)$ state, observed by CDF in $B^+\to K^+\phi J/\psi$ decays \cite{cdfx4140},
can correspond in our model to the scalar $0^{++}$ charmed-strange
diquark-antidiquark 
$[cs]_{S=1}[\bar c \bar s]_{S=1}$ ground state, which predicted mass is
4110~MeV, or the axial vector $1^{++}$ one $([cs]_{S=0}[\bar c \bar
s]_{S=1}+[cs]_{S=1}[\bar c \bar s]_{S=0})/\sqrt2)$ with calculated
mass 4113~MeV \cite{tetr,htetr}.    

Two of the three charmonium-like charged $X^\pm$ states reported by
Belle \cite{bellexpl}, which are explicitly exotic, 
can be interpreted in our model as tetraquark states. We do not have tetraquark
candidates for  the
$X(4040)^+$ structure, while the $X(4250)^+$ can be considered 
as the charged partner of 
the $1^{-}$ $1P$ state $[cu]_{S=0}[\bar c \bar d]_{S=0}$ or as  the $0^-$  $1P$ state of
the $([cu]_{S=0}[\bar c \bar
d]_{S=1}+[cu]_{S=1}[\bar c \bar d]_{S=0})/\sqrt2)$ tetraquark with
predicted masses 4244~MeV and 4267~MeV, respectively \cite{htetr}. The $X(4430)^+$
could be the first radial ($2S$)
excitation of the $1^+$  $X(3872)$ tetraquark or  the $0^{+}$ $2S$
$[cu]_{S=1}[\bar c \bar d]_{S=1}$ tetraquark,  which have  very
close masses 4431~MeV and 4434~MeV \cite{htetr}. 

Very recently the Belle Collaboration \cite{bellezb} reported the
observation of two charged bottomonium-like resonances, the
$Z_b(10610)$ and $Z_b(10650)$, in the
$\pi^\pm\Upsilon$ and $\pi^\pm h_b$ mass spectra close to the
open-bottom ($B\bar B^*$ and $B^*\bar B^*$) production
thresholds. The analysis of the charged pion angular 
distributions favours a $J^P=1^+$ spin-parity assignment for both
states \cite{bellezb}.  In this mass region we do not have any bottom
diquark-antidiquark 
tetraquarks with such quantum numbers \cite{htetr}. The possible
interpretations of these exotic bottomonium-like states are discussed in
Ref.~\cite{ali}. 

As we see, a consistent picture of the excited quarkonium states
emerges in our model. All well-established states and most of the states, which
need additional experimental confirmation, can be interpreted  as
excited quarkonium  or diquark-antidiquark tetraquark states.

\section{Conclusions}
\label{sec:concl}

The mass spectra of charmonia, bottomonia and $B_c$ mesons were
calculated in the framework of the QCD-motivated relativistic quark model based on
the quasipotential approach. Highly radially and orbitally excited quarkonium
states were considered. To achieve this goal, we treated the dynamics
of heavy quarks in quarkonia completely relativistically without
application of the nonrelativistic $v^2/c^2$ expansion. The known
one-loop radiative corrections were also taken into account in order
to improve the agreement with experiment. The
comparison of new results with the previous consideration within the
$v^2/c^2$ expansion \cite{efg} indicates that relativistic effects
become significant with the increase of the excitation and are
particularly important for charmonium.    

On this basis, the Regge trajectories of heavy quarkonia were
constructed both in the ($J, M^2$) and $(n_r,M^2)$ planes. A
different behaviour of these trajectories was observed for parent and
daughter trajectories. All daughter trajectories turn out to be almost
linear and parallel, while parent trajectories exhibit some
nonlinearity. Such nonlinearity occurs only in the vicinity of ground
states and few lowest excitations and is mostly pronounced for
bottomonia. For charmonia this nonlinearity is only marginal, and its
account does not significantly improve the fit. It was shown that the
masses of the excited states of heavy quarkonia are determined by
the average distances between quarks larger than 0.5~fm, where the
linear confining part of the quark-antiquark interaction
dominates. This leads to the emergence of almost linear Regge
trajectories. On the other hand, a few lowest states have average sizes
smaller than 0.5~fm and fall in the region, where both the Coulomb and confining
potentials play an important role. As a result, the parent Regge
trajectories exhibit a certain nonlinearity in this region. The parameters
(slopes, intercepts and nonlinearity) of
both linear and nonlinear Regge trajectories were determined. They
were compared to the slopes of the linear Regge trajectories of light
\cite{lregge} and heavy-light \cite{hlrt} mesons calculated
previously. It was found that the slope of the meson Regge trajectory
is mainly determined by the mass of the heaviest quark $m_Q$.

A detailed comparison of the calculated heavy quarkonium masses with
available experimental data  was carried out. It was found
that all data for the states below open flavour production threshold
are well reproduced in our model: the difference between predicted and
measured masses does not exceed a few MeV.  For higher excited states,
which are above this threshold, most of the well-established
conventional  states are also well described by our approach, the 
difference between theory and experiment being somewhat larger, but still
within 20 MeV. It was shown that these states fit well to the
corresponding Regge trajectories. Other states, which have unexpected
properties and are therefore believed to have an exotic origin, were also
discussed. As it was shown in our previous calculation \cite{tetr,htetr},
most of these states can be described as  diquark-antidiquark
tetraquarks. Therefore we have a self-consistent picture of the heavy
quarkonium spectra. Future experimental studies of
the yet unobserved conventional quarkonium states and a clarification of 
the nature and quantum numbers of the exotic quarkonium-like states
will provide a further test of our model.            

%\vspace*{1cm}

\acknowledgements
The authors are grateful to M. M\"uller-Preussker,   
V. Matveev, V. Savrin and M. Wagner for support and discussions. One of us
(V.O.G.) thanks  the particle theory group at Humboldt University for kind hospitality. 
V.O.G. gratefully acknowledges the financial support  by 
Deutscher Akademischer Austauschdienst (DAAD).

\end{document}